\documentclass{article}
\usepackage[letterpaper]{geometry}
\usepackage[T1]{fontenc}
\usepackage[utf8]{inputenc}
\usepackage[english]{babel}
\usepackage{csquotes}
\usepackage{latexsym, amsmath,amssymb}
\usepackage{graphicx}
\usepackage{subcaption}
\usepackage{booktabs,multicol}
\usepackage[table]{xcolor}
\usepackage{comment}
\usepackage[normalem]{ulem}

\usepackage{authblk}

\usepackage[hyphens]{url}
\usepackage[breaklinks=true,linkcolor=blue, citecolor=blue, urlcolor=blue, colorlinks=true]{hyperref}

\usepackage[
    backend=biber,
    style=numeric,
    citestyle=numeric-comp,
    sorting=none,
    bibencoding=UTF-8,
    giveninits=true,
    maxbibnames=1000,
]{biblatex}
\usepackage[textsize=footnotesize,textwidth=2cm]{todonotes}

\usepackage[xcolor]{changebar}
\newcommand{\delete}[1]{\sloppy\cbcolor{red}\textcolor{red}{\cbdelete \sout{#1}}}
\newcommand{\deletenoso}[1]{{\textcolor{red}{#1}}}    

\renewcommand{\delete}[1]{}
\renewcommand{\deletenoso}[1]{}

\newcommand\joss{\textit{JOSS}}

\title{Journal of Open Source Software (JOSS): design and first-year review}

\author[1]{Arfon M.~Smith\thanks{Corresponding author, \href{mailto:arfon@stsci.edu}{arfon@stsci.edu}}}
\author[2]{Kyle E.~Niemeyer}
\author[3]{Daniel S.~Katz}
\author[4]{Lorena A.~Barba}
\author[5]{George~Githinji}
\author[6]{Melissa Gymrek}
\author[7]{Kathryn D.~Huff}
\author[8]{Christopher R.~Madan}
\author[9]{Abigail Cabunoc Mayes}
\author[10]{Kevin M.~Moerman}
\author[11]{Pjotr Prins}
\author[12]{Karthik Ram}
\author[13]{Ariel Rokem}
\author[14]{Tracy K.~Teal}
\author[15]{Roman Valls Guimera}
\author[13]{Jacob~T.~Vanderplas}

\date{January 2018}

\affil[1]{Data Science Mission Office, Space Telescope Science Institute, Baltimore, MD, USA}
\affil[2]{School of Mechanical, Industrial, and Manufacturing Engineering, Oregon State University, Corvallis, OR, USA}
\affil[3]{National Center for Supercomputing Applications \& Department of Computer Science \& Department of Electrical and Computer Engineering \& School of Information Sciences, University of Illinois at Urbana--Champaign, Urbana, IL, USA}
\affil[4]{Department of Mechanical \& Aerospace Engineering, George Washington University, Washington, DC, USA}
\affil[5]{KEMRI--Wellcome Trust Research Programme, Kilifi, Kenya}
\affil[6]{Departments of Medicine \& Computer Science and Engineering, University of California, San Diego, CA, USA}
\affil[7]{Department of Nuclear, Plasma, and Radiological Engineering, University of Illinois at Urbana--Champaign, Urbana, IL, USA}
\affil[8]{School of Psychology, University of Nottingham, Nottingham, United Kingdom}
\affil[9]{Mozilla Foundation, Toronto, Ontario, Canada}
\affil[10]{MIT Media Lab, Massachusetts Institute of Technology, Cambridge, MA, USA \& Trinity Centre for Bioengineering, Trinity College, The University of Dublin, Dublin, Ireland}
\affil[11]{University of Tennessee Health Science Center, Memphis, TN, USA \& University Medical Centre Utrecht, Utrecht, The Netherlands}
\affil[12]{Berkeley Institute for Data Science, University of California, Berkeley, Berkeley, CA, USA}
\affil[13]{eScience Institute, University of Washington, Seattle, WA, USA}
\affil[14]{Data Carpentry, Davis, CA, USA}
\affil[15]{University of Melbourne Centre for Cancer Research, University of
Melbourne, Melbourne, Australia}

\begin{document}

\maketitle


\begin{abstract}
This article describes the motivation, design, and progress of the Journal of Open Source Software (\joss{}).
\joss{} is a free and open-access journal that publishes articles describing research software.
It has the dual goals of improving the quality of the software submitted and providing a mechanism for research software developers to receive credit.
While designed to work within the current merit system of science, \joss{}
addresses the dearth of rewards for key contributions to science made in the form of software.
\joss{} publishes articles that encapsulate scholarship contained in the software itself, and its rigorous peer review targets the software components: functionality, documentation, tests, continuous integration, and the license.
A \joss{} article contains an abstract describing the purpose and functionality of the software, references, and a link to the software archive.
The article is the entry point of a \joss{} submission, which encompasses the full set of software artifacts.
Submission and review proceed in the open, on GitHub.
Editors, reviewers, and authors work collaboratively and openly.
Unlike other journals, \joss{} does not reject articles requiring major revision; while not yet accepted, articles remain visible and under review until the authors make adequate changes (or withdraw, if unable to meet requirements).
Once an article is accepted, \joss{} gives it a digital object identifier (DOI), deposits its metadata in Crossref, and the article can begin collecting citations on indexers like Google Scholar and other services.
Authors retain copyright of their \joss{} article, releasing it under a Creative Commons Attribution 4.0 International License.
In its first year, starting in May 2016, \joss{} published 111 articles, with more than 40 additional articles under review.
\joss{} is a sponsored project of the nonprofit organization NumFOCUS and is an affiliate of the Open Source Initiative (OSI).
\end{abstract}

\section{Introduction}

Modern scientific research produces many outputs beyond traditional articles and books. Among these, research software is critically important for a broad spectrum of fields.
Current practices for publishing and citation do not, however, acknowledge software as a first-class research output. This deficiency means that researchers who develop software face critical career barriers.
The \textit{Journal of Open Source Software} (\joss{}) was founded in May 2016 to offer a solution within the existing publishing mechanisms of science.
It is a developer-friendly, free and open-access, peer-reviewed journal for research software packages.
By its first anniversary, \joss{} had published more than a hundred articles.
This article discusses the motivation for creating a new software journal, delineates the editorial and review process, and summarizes the journal's first year of operation via submission statistics.
We expect this article to be of interest to three core audiences: (1) researchers who develop software and could submit their work to \joss{},
(2) those in the community with an interest in advancing scholarly communications who may appreciate the technical details of the \joss{} journal framework, and
(3) those interested in possibilities for citing software in their own research publications.

The sixteen authors of this article are the members of the \joss{} Editorial Board at the end of its first year (May 2017). Arfon Smith is the founding editor-in-chief, and the founding editors are Lorena A.~Barba, Kathryn Huff, Daniel Katz, Christopher Madan, Abigail Cabunoc Mayes, Kevin Moerman, Kyle Niemeyer, Karthik Ram, Tracy Teal, and Jake Vanderplas.
Five new editors joined in the first year to handle areas not well covered by the original editors, and to help manage the large and growing number of submissions. They are George Githinji, Melissa Gymrek, Pjotr Prins, Ariel Rokem, and Roman Valls Guimera.
(Since then, we added three more editors: Jason Clark, Lindsey Heagy, and Thomas Leeper.)

The \joss{} editors are firm supporters of open-source software for research, with extensive knowledge of the practices and ethics of open source. This knowledge is reflected in the \joss{} submission system, peer-review process, and infrastructure. The journal offers a familiar environment for developers and authors to interact with reviewers and editors, leading to a citable published work: a software article.
The article describes the software at a high level, and the software itself includes both source code and associated artifacts such as tests, documentation, and examples.
With a Crossref digital object identifier (DOI), the article is able to collect citations, empowering the developers/authors to gain career credit for their work.
\joss{} thus fills a pressing need for computational researchers to advance professionally, while promoting higher quality software for science.
\joss{} also supports the broader open-science movement by encouraging researchers to share their software openly and follow best practices in its development.

\section{Background and motivation}\label{background}


A 2014 study of UK Russell Group Universities~\cite{Hettrick} reports that $\sim$90\% of academics surveyed said they use software in their research, while more than 70\% said their research would be impractical without it.
About half of these UK academics said they develop their own software while in the course of doing research.
Similarly, a 2017 survey of members of the US National Postdoctoral Association found that 95\% used research software, and 63\% said their research would be impractical without it~\cite{US-PDA-survey}.

Despite being a critical part of modern research, software lacks support across the scholarly ecosystem for its publication, acknowledgement, and citation~\cite{Niemeyer:2016sc}.
Academic publishing has not changed substantially since its inception.
Science, engineering, and many other academic fields still view research articles as the key indicator of research productivity, with research grants being another important indicator.
Yet, the research article is inadequate to fully describe modern, data-intensive, computational research.
\joss{} focuses on research software and its place in the scholarly publishing ecosystem.

\subsection{Why publish software?}

Most academic fields still rely on a one-dimensional credit model where academic articles and their associated citations are the dominant factor in the success of a researcher's career. Software creators, in order to increase the likelihood of receiving career credit for their work, often choose to publish ``software articles'' that act as placeholder publications pointing to their software.
At the same time, recent years have seen a push for sharing open research software~\cite{Barnes:2010ut,Vandewalle:2012cl,Morin:2012hz,Ince:2012iy,NatureMethodsEditorialBoard:2014gu,Prins:natbio}.

Beyond career-credit arguments for software creators, publishing research software enriches the scholarly record. Buckheit and Donoho paraphrased Jon Claerbout, a pioneer of reproducible research, as saying: ``An article about a computational result is advertising, not scholarship. The actual scholarship is the full software environment, code and data, that produced the result.''~\cite{Buckheit1995}.
The argument that articles about computational science are not satisfactory descriptions of the work, needing to be supplemented by code and data, is more than twenty years old! Yet, despite the significance of software in modern research, documenting its use and including it in the scholarly ecosystem presents numerous challenges.

\subsection{Challenges of publishing software}

The conventional publishing mechanism of science is the research article, and a researcher's career progression hinges on collecting citations for published works.
Unfortunately, software citation~\cite{Smith2016} is in its infancy (as is data citation~\cite{data-citation,10.7717/peerj-cs.1}).
Publishing the software itself and receiving citation credit for it may be a better long-term solution, but this is still impractical.
Even when software (and data) are published so that they can be cited, we do not have a standard culture of peer review for them. This leads many developers today to publish software articles.

The developer's next dilemma is where to publish, given the research content, novelty, length and other features of a software article.
Since 2012, Neil Chue Hong has maintained a growing list of journals that accept software articles~\cite{software-papers-list}.
He includes both generalist journals, accepting software articles from a variety of fields, and domain-specific journals, accepting both research and software articles in a given field.
For many journals, particularly the domain-specific ones, a software article must include novel results to justify publication.

From the developer's point of view, writing a software article can involve a great deal of extra work. Good software includes documentation for both users and developers that is sufficient to make it understandable. A software article may contain much of the same content, merely in a different format, and developers may not find value in rewriting their documentation in a manner less useful to their users and collaborators.
These issues may lead developers to shun the idea of software articles and prefer to publish the software itself. Yet, software citation is not common and the mostly one-dimensional credit model of academia (based on article citations) means that publishing software often does not ``count'' for career progression~\cite{Smith2016,Niemeyer:2016sc}.

\section{The Journal of Open Source Software}

To tackle the challenges mentioned above, the \textit{Journal of Open Source Software} (\joss{}) launched in May 2016~\cite{arfondotorg} with the goal of drastically reducing the overhead of publishing software articles.
\joss{} offers developers a venue to publish their complete research software wrapped in relatively short high-level articles, thus enabling citation credit for their work.
In this section we describe the goals and principles, infrastructure, and business model of \joss{}, and compare it with other software journals.

\subsection{Goals and principles}

\joss{} articles are deliberately short and only include an abstract describing the high-level functionality of the software, a list of the authors of the software (with their affiliations), a list of key references, and a link to the software archive and software repository. Articles are not allowed to include other content often found in software articles, such as descriptions of the API (application programming interface) and novel research results obtained using the software.
The software API should already be described in the software documentation, and domain research results do not belong in \joss{}---these should be published in a domain journal.
Unlike most journals, which ease discoverability of new research and findings, \joss{} serves primarily as a mechanism for software developers\slash authors to improve and publish their research software.
Thus, software discovery is a secondary feature.

The \joss{} design and 
implementation are based on the following principles:
\begin{itemize}
    \item Other than their short length, \joss{} articles are conventional articles in every other sense: the journal has an ISSN, articles receive Crossref DOIs with high-quality submission metadata, and articles are appropriately archived.
    \item Because software articles are ``advertising'' and simply pointers to the \textit{actual} scholarship (the software), short abstract-length submissions are sufficient for these ``advertisements.''
    \item Software is a core product of research and therefore the software itself should be archived appropriately when submitted to and reviewed in \joss{}.
    \item Code review, documentation, and contributing guidelines are important for open-source software and should be part of any review.
    In \joss{}, they are the focus of peer review.
    (While a range of other journals publish software,
    with various peer-review processes, the focus of the review is usually the submitted article and reviewers might not even look at the code.)
    The \joss{} review process itself, described in \S\ref{thereview}, was based on the on-boarding checklist for projects joining the rOpenSci collaboration~\cite{ropensci}.
\end{itemize}

Acceptable \joss{} submissions also need to meet the following criteria:

\begin{itemize}
    \item The software must be open source by the Open Source Initiative (OSI) definition (\href{https://opensource.org}{opensource.org}).
    \item The software must have a research application.
    \item The submitter should be a major contributor to the software they are submitting.
    \item The software should be a significant new contribution to the available open-source software that either enables some new research challenge(s) to be addressed or makes addressing research challenges significantly better (e.g., faster, easier, simpler).
    \item The software should be feature-complete, i.e., it cannot be a partial solution.
\end{itemize}

\subsection{How \joss{} works}\label{howitworks}

\joss{} is designed as a small collection of open-source tools that leverage existing infrastructure such as GitHub, Zenodo, and Figshare. A goal when building the journal was to minimize the development of new tools where possible.

\subsubsection*{The \joss{} web application and submission tool}

\begin{figure}[t]
\centering
\includegraphics[width=1.0\textwidth]{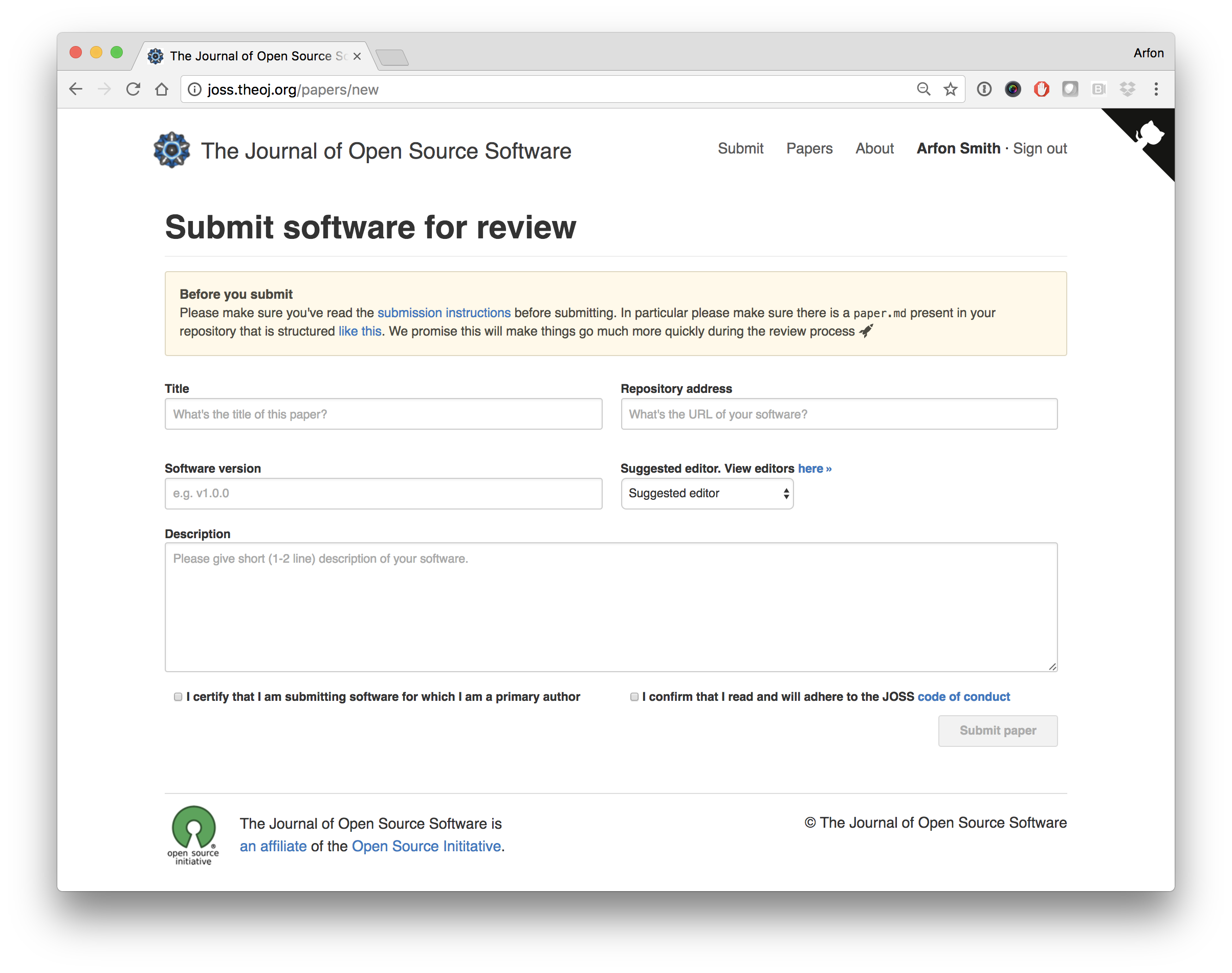}
\caption{The \joss{} submission page. A minimal amount of information is required for new submissions.
\label{fig:submission}}
\end{figure}

The \joss{} web application and submission tool is hosted at \href{http://joss.theoj.org}{http://joss.theoj.org}. It is a simple Ruby on Rails web application~\cite{joss-site} that lists accepted articles, provides the article submission form (see Figure~\ref{fig:submission}), and hosts journal documentation such as author submission guidelines.
This application also automatically creates the review issue on GitHub once a submission has been pre-reviewed by an editor and accepted to start peer review in \joss{}.

\subsubsection*{Open peer review on GitHub}

\begin{figure}[t]
\centering
\includegraphics[width=1.0\textwidth]{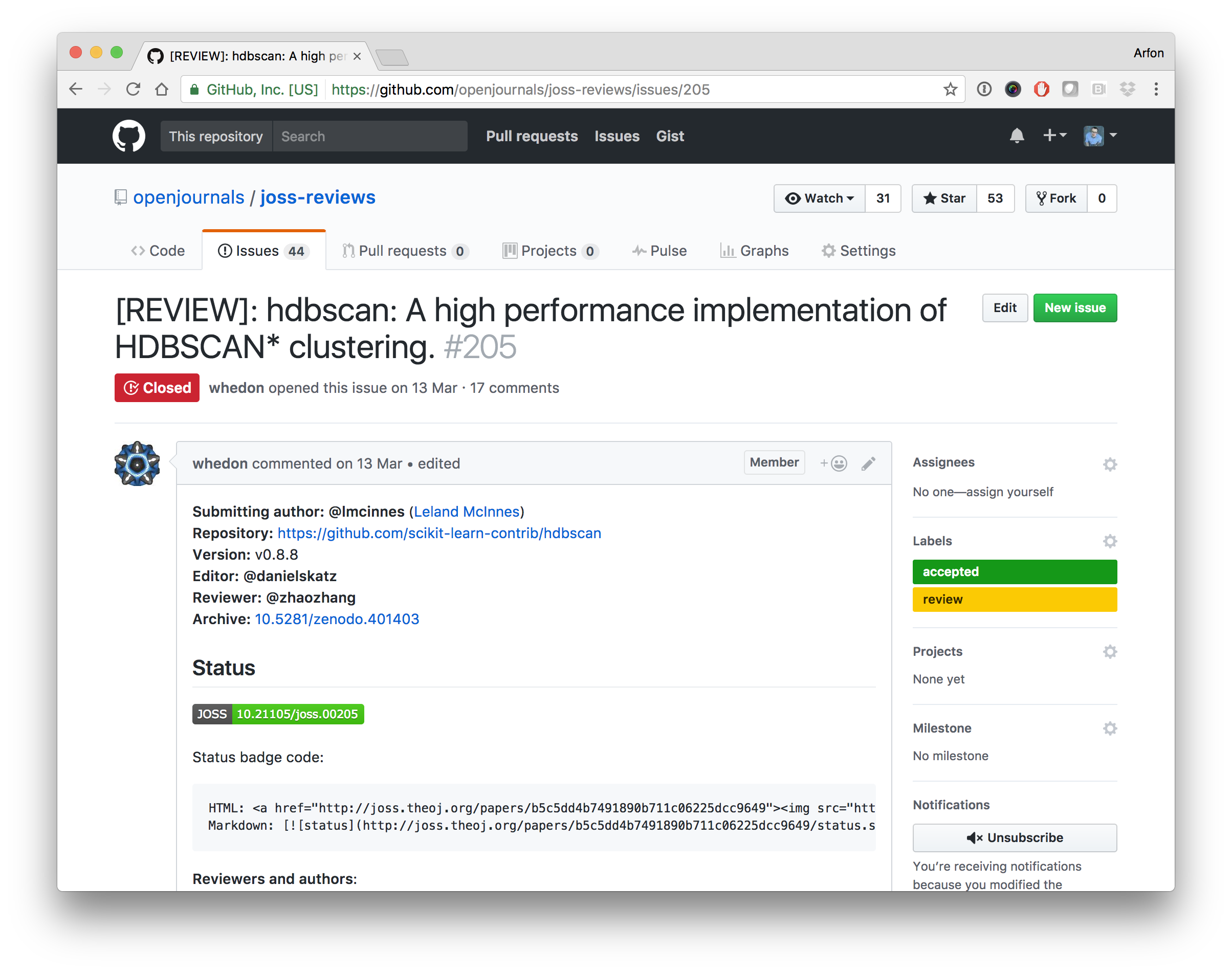}
\caption{The \texttt{hdbscan} GitHub review issue.
\label{fig:review}}
\end{figure}

\joss{} conducts reviews on the \texttt{joss-reviews} GitHub repository~\cite{joss-reviews}.  Review of a submission begins with the opening of a new GitHub issue, where the editor-in-chief assigns an editor, the editor assigns a reviewer, and interactions between authors, reviewer(s), and editor proceed in the open. Figure~\ref{fig:review} shows an example of a recent review for the (accepted) \texttt{hdbscan} package~\cite{McInnes2017}.
The actual review includes the code, software functionality\slash performance claims, test suite (if present), documentation, and any other material associated with the software.

\subsubsection*{Whedon and the Whedon-API}

Many of the tasks associated with \joss{} reviews and editorial management are automated.
A core RubyGem library named \texttt{Whedon}~\cite{whedon-gem} handles common tasks associated with managing the submitted manuscript, such as compiling the article (from its Markdown source) and creating Crossref metadata.
An automated bot, \texttt{Whedon-API}~\cite{whedon-api}, handles other parts of the review process (such as assigning editors and reviewers based on editor input) and leverages the \texttt{Whedon} RubyGem library.
For example, to assign the editor for a submission, one may type the following command in a comment box within the GitHub issue: \texttt{@whedon assign @danielskatz as editor}. Similarly, to assign a reviewer, one enters: \texttt{@whedon assign @zhaozhang as reviewer} (where the reviewer and editor GitHub handles identify them). The next section describes the review process in more detail.

\subsection{Business model and content licensing}

\joss{} is designed to run at minimal cost with volunteer labor from editors and reviewers. The following fixed costs are currently incurred:

\begin{itemize}
  \item{Crossref membership: \$275. This is a yearly fixed cost for the \joss{} parent entity---\textit{Open Journals}---so that article DOIs can be registered with Crossref.}
  \item{Crossref article DOIs: \$1. This is a fixed cost per article.}
  \item{\joss{} web application hosting (currently with Heroku): \$19 per month}
\end{itemize}

Assuming a publication rate of 100 articles per year results in a core operating cost of $\sim$\$6 per article.
With 200 articles per year---which seems possible for the second
year---the cost drops to $\sim$\$3.50 per article:
\begin{align}\label{costs}
(\$275 + (\$1 \times 100) + (\$19 \times 12)) / 100 &= \$6.03 \\
(\$275 + (\$1 \times 200) + (\$19 \times 12)) / 200 &= \$3.51 \;.
\end{align}

Submitting authors retain copyright of \joss{} articles and accepted articles are published under a Creative Commons Attribution 4.0 International License~\cite{cc}.
Any code snippets included in \joss{} articles are subject to the MIT license~\cite{mit} regardless of the license of the submitted software package under review, which itself must be licensed under an OSI-approved license (see \href{https://opensource.org/licenses/alphabetical}{opensource.org/licenses/alphabetical} for a complete list).

\subsection{Comparison with other software journals}
\label{comparison}

A good number of journals now accept, review, and publish software articles~\cite{software-papers-list},
which we group into two categories.
The first category of journals include those similar to \joss{}, which do not focus on a specific domain and only consider submissions of software\slash software articles:
the \textit{Journal of Open Research Software} (\textit{JORS},
\href{http://openresearchsoftware.metajnl.com}{openresearchsoftware.metajnl.com}),
\textit{SoftwareX}
(\href{https://www.journals.elsevier.com/softwarex/}{journals.elsevier.com/softwarex/}), and now \joss{}.
Both \textit{JORS}~\cite{jorsreview} and \textit{SoftwareX}~\cite{els-software}
now review both the article text and the software.
In \joss{}, the review process focuses mainly on the software and associated material (e.g., documentation) and less on the article text, which is intended to be a brief description of the software.
The role and form of peer review also varies across journals.
In \textit{SoftwareX} and \textit{JORS}, the goal of the review is both to decide if the article is acceptable for publication and to improve it iteratively through a non-public, editor-mediated interaction between the authors and the anonymous reviewers.
In contrast, \joss{} has the goal of accepting most articles after improving them as needed, with the reviewers and authors communicating directly and publicly through GitHub issues.

The second category includes domain-specific journals that either accept software articles as a special submission type or exclusively consider software articles targeted at the domain.
For example, \textit{Collected Algorithms} (CALGO, \href{http://www.acm.org/calgo/}{acm.org/calgo}) is a long-running venue for reviewing and sharing mathematical algorithms associated with articles published in \textit{Transactions on Mathematical Software} and other ACM journals.
However, CALGO authors must transfer copyright to ACM and software is not available under an open-source license---this contrasts with \joss{}, where authors retain copyright and software must be shared under an open-source license.
\textit{Computer Physics Communications} (\href{https://www.journals.elsevier.com/computer-physics-communications}{journals.elsevier.com/computer-physics-communications}) and \textit{Geoscientific Model Development} (\href{https://www.geoscientific-model-development.net/}{geoscientific-model-development.net}) publish full-length articles describing application software in computational physics and geoscience, respectively, where review primarily focuses on the article.
Chue Hong maintains a list of journals in both categories~\cite{software-papers-list}.

\section{Peer review in \joss{}}
\label{thereview}

In this section, we illustrate the \joss{} submission and review process using a representative example, document the review criteria provided to authors and reviewers, and explain a fast-track option for already-reviewed rOpenSci contributions.

\subsection{The \joss{} process}

\begin{figure}[htp]
\centering
\includegraphics[width=0.75\textwidth]{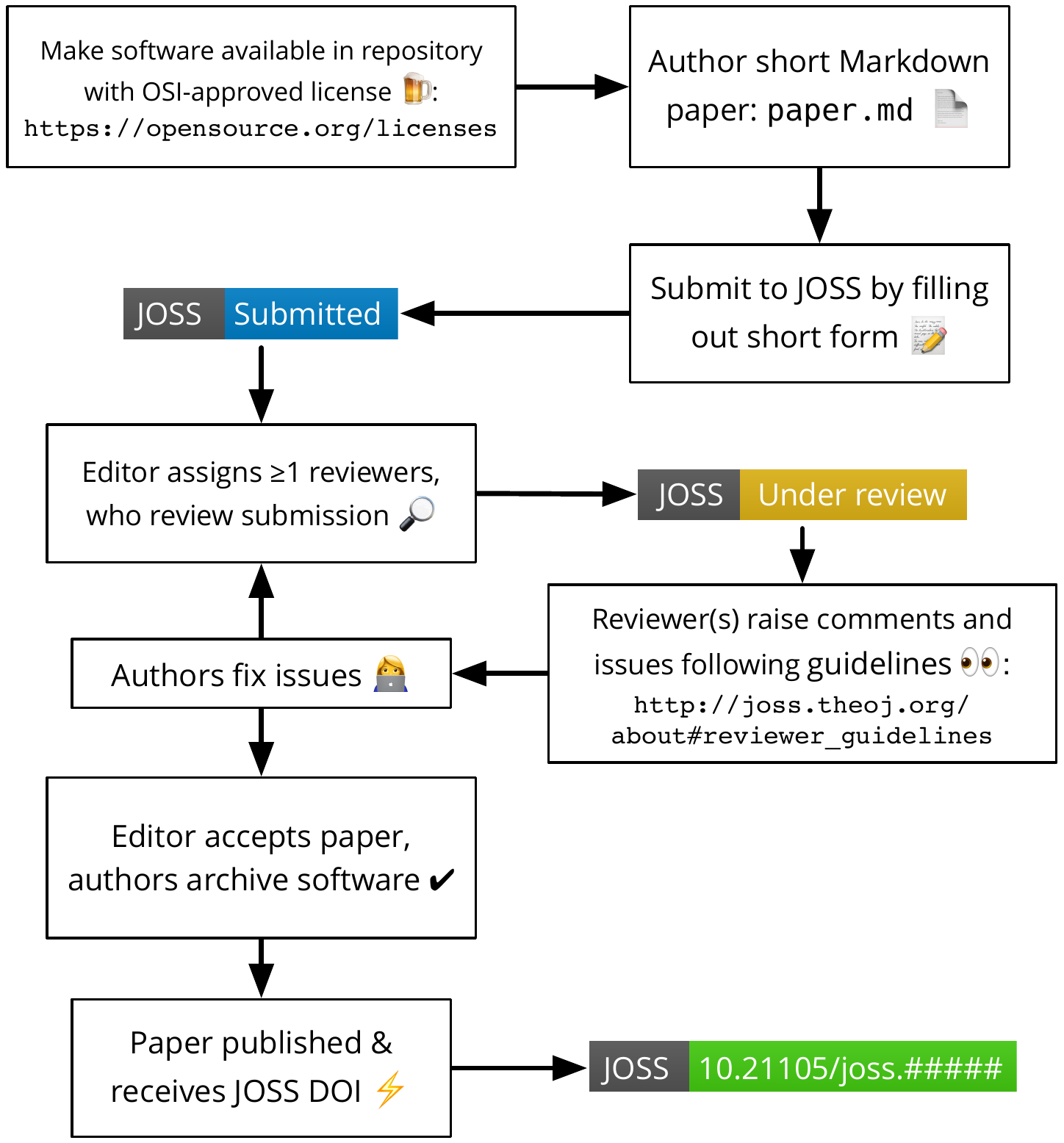}
\caption{The \joss{} submission and review flow including the various status badges that can be embedded on third-party settings such as GitHub README documentation~\cite{JOSS-publication-workflow}.
\label{fig:submission-flow}}
\end{figure}

Figure~\ref{fig:submission-flow} shows a typical \joss{} submission and review process, described here in more detail using the \texttt{hdbscan} package~\cite{McInnes2017} as an example:

\begin{enumerate}
\item Leland McInnes submitted the \texttt{hdbscan} software and article to \joss{} on 26 February 2017 using the web application and submission tool.
The article is a Markdown file named \texttt{paper.md}, visibly located in the software repository (here, and in many cases, placed together with auxiliary files in a \texttt{paper} directory).

\item Following a routine check by a \joss{} administrator, a ``pre-review'' issue was created in the \texttt{joss-reviews} GitHub repository~\cite{hdbscan-joss-pre-review}. In this pre-review issue, an editor (Daniel S.~Katz) was assigned, who then identified and assigned a suitable reviewer (Zhao Zhang).
Editors generally identify one or more reviewers from a pool of volunteers based on provided programming language and\slash or domain expertise.\footnote{Potential reviewers can volunteer via \url{http://joss.theoj.org/reviewer-signup.html}}

The editor then asked the automated bot \texttt{Whedon} to create the main submission review issue via the command \texttt{@whedon start review magic-word=bananas}. (``\texttt{magic-word=bananas}'' is a safeguard against accidentally creating a review issue prematurely.)

\item The reviewer then conducted the submission review~\cite{hdbscan-joss-review} (see Figure~\ref{fig:review}) by working through a checklist of review items, as described in \S\ref{review-details}. The author, reviewer, and editor discussed any questions that arose during the review, and once the reviewer completed their checks, they notified the submitting author and editor.
Compared with traditional journals, \joss{} offers the unique feature of holding a discussion---in the open within a GitHub issue---between the reviewer(s), author(s), and editor.
Like a true conversation, discussion can go back and forth in minutes or seconds, with all parties contributing at will. This contrasts with traditional journal reviews, where the process is merely an exchange between the reviewer(s) and author(s), via the editor, which can take months for each communication, and in practice is limited to one or two, perhaps three in some cases, exchanges due to that delay~\cite{tennant-peerreview}.

Note that \joss{} reviews are subject to a code of conduct~\cite{code-of-conduct}, adopted from the Contributor Covenant Code of Conduct~\cite{contributor-covenant-coc}.
Both authors and reviewers must confirm that they have read and will adhere to this Code of Conduct, during submission and with their review, respectively.

\item After the review was complete, the editor asked the submitting author to make a permanent archive of the software (including any changes made during review) with a service such as Zenodo or Figshare, and to post a link to the archive in the review thread. This link, in the form of a DOI, was associated with the submission via the command \texttt{@whedon set 10.5281/zenodo.401403 as archive}.

\item The editor-in-chief used the \texttt{Whedon} RubyGem library on his local machine to produce the compiled PDF, update the \joss{} website, deposit Crossref metadata, and issue a DOI for the submission (\href{https://doi.org/10.21105/joss.00205}{10.21105/joss.00205}).

\item Finally, the editor-in-chief updated the review issue with the \joss{} article DOI and closed the review. The submission was then accepted into the journal.

\end{enumerate}

Authors can also first submit a pre-submission inquiry via an issue in the main \joss{} repository~\cite{joss-site} if they have questions regarding the suitability of their software for publication, or for any other questions.

\subsection{\joss{} review criteria}\label{review-details}

As previously mentioned, the \joss{} review is primarily concerned with the material
in the software repository, focusing on the software and documentation.
We do not ask authors to use their software in a research study or include research results in their article beyond as examples; submissions focused on results rather than software should be submitted to research journals.
The specific items in the reviewer checklist are:

\begin{itemize}
\item Conflict of interest

\begin{itemize}
    \item As the reviewer I confirm that
    I have read the \joss{} \href{https://github.com/openjournals/joss/blob/master/COI.md}{conflict of interest policy}
    and that there are no conflicts of interest for me to review this work.
\end{itemize}

\item Code of Conduct
\begin{itemize}
    \item I confirm that I read and will adhere to the \href{http://joss.theoj.org/about#code_of_conduct}{\joss{} code of conduct}.
\end{itemize}

\item General checks

\begin{itemize}
    \item \textbf{Repository}: Is the source code for this software available at the repository URL?
    \item \textbf{License}: Does the repository contain a plain-text LICENSE file with the contents of an OSI-approved software license?
    \item \textbf{Version}: Does the release version given match the GitHub release?
    \item \textbf{Authorship}: Has the submitting author made major contributions to the software?
\end{itemize}

\item Functionality

\begin{itemize}
    \item \textbf{Installation}: Does installation proceed as outlined in the documentation?
    \item \textbf{Functionality}: Have the functional claims of the software been confirmed?
    \item \textbf{Performance}: Have any performance claims of the software been confirmed?
\end{itemize}

\item Documentation

\begin{itemize}
    \item \textbf{A statement of need}: Do the authors clearly state what problems the software is designed to solve and who the target audience is?
    \item \textbf{Installation instructions}: Is there a clearly-stated list of dependencies? Ideally these should be handled with an automated package management solution.
    \item \textbf{Example usage}: Do the authors include examples of how to use the software (ideally to solve real-world analysis problems)?
    \item \textbf{Functionality documentation}: Is the core functionality of the software documented to a satisfactory level (e.g., API method documentation)?
    \item \textbf{Automated tests}: Are there automated tests or manual steps described so that the function of the software can be verified?
    \item \textbf{Community guidelines}: Are there clear guidelines for third parties wishing to 1) contribute to the software, 2) report issues or problems with the software, and 3) seek support?
\end{itemize}

\item Software paper

\begin{itemize}
    \item \textbf{Authors}: Does the \texttt{paper.md} file include a list of authors with their affiliations?
    \item \textbf{A statement of need}: Do the authors clearly state what problems the software is designed to solve and who the target audience is?
    \item \textbf{References}: Do all archival references that should have a DOI list one (e.g., papers, datasets, software)?
\end{itemize}

\end{itemize}

\subsection{Fast track for reviewed rOpenSci contributions}

For submissions of software that has already been reviewed under rOpenSci's rigorous
onboarding guidelines~\cite{Ram:2016ws,Ram2017}, \joss{} does not perform
further review. The editor-in-chief is alerted with a note
``This submission has been accepted to rOpenSci. The review thread can be
found at \texttt{[LINK TO ONBOARDING ISSUE]},''  allowing such submissions to be fast-tracked to  acceptance.

\section{A review of the first year}\label{firstyear}

\begin{figure}[htbp]
    \centering
    \begin{subfigure}[b]{0.7\textwidth}
        \includegraphics[width=\textwidth]{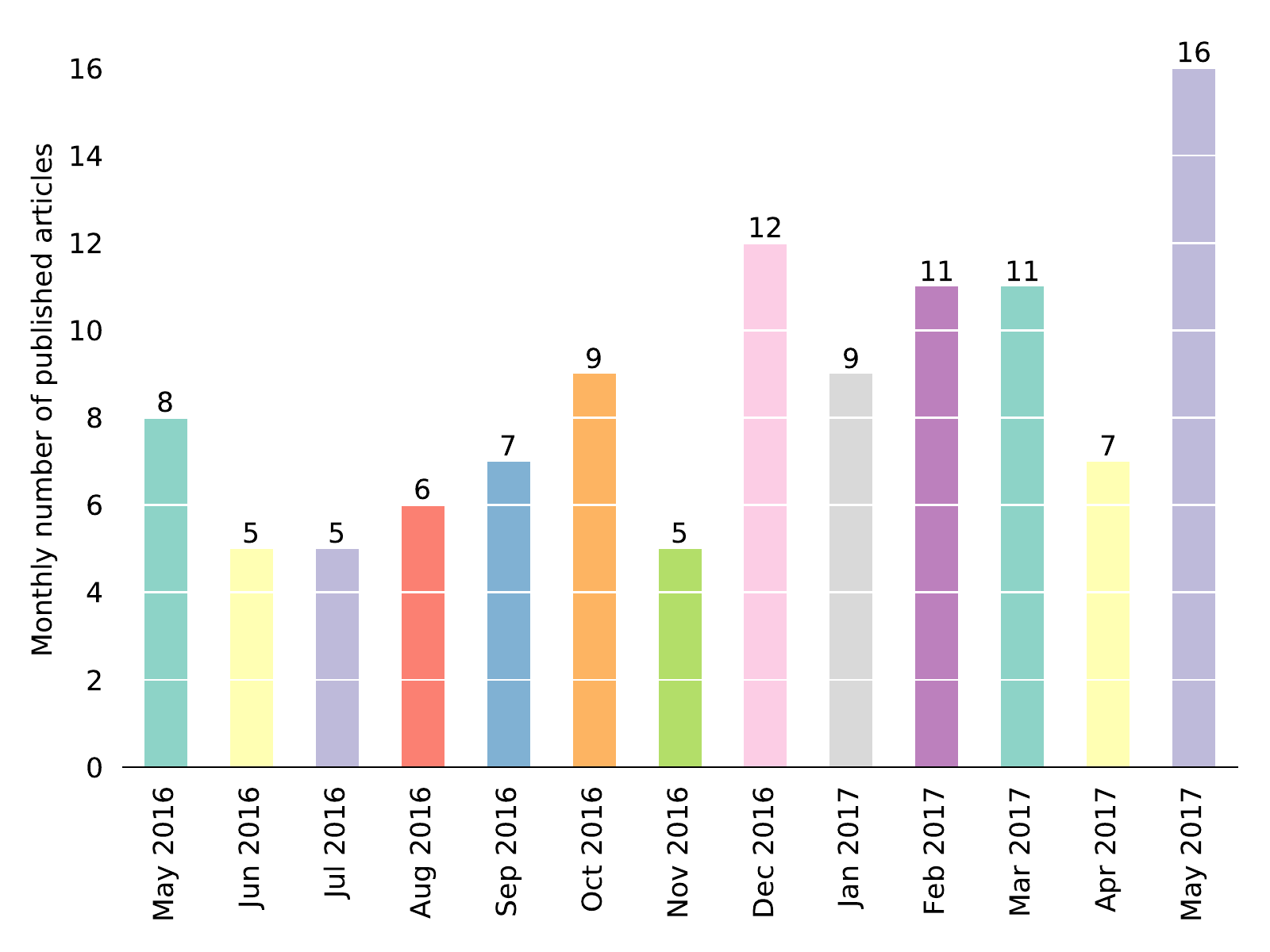}
        \caption{Numbers of articles published per month.}
    \end{subfigure}
    \\
    \begin{subfigure}[b]{0.7\textwidth}
        \includegraphics[width=\textwidth]{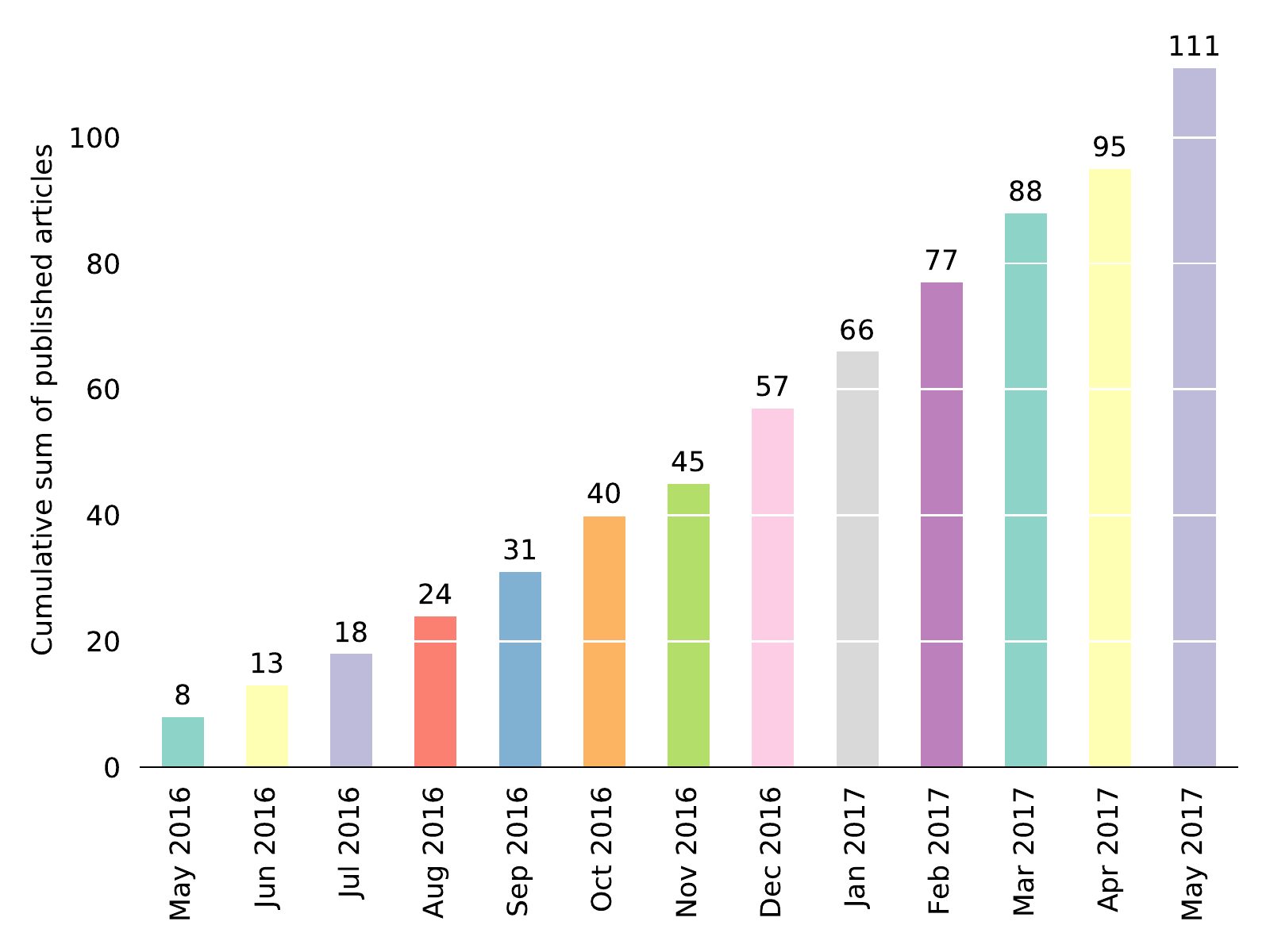}
        \caption{Cumulative sum of numbers of articles published per month.}
    \end{subfigure}
    \caption{Statistics of articles published in \joss{} since its inception in May 2016 through May 2017.
    Data, plotting script, and figure files are available~\cite{JOSS-data-figs}.}
    \label{fig:article_stats}
\end{figure}

By the end of May 2017, \joss{} published 111 articles since its
inception in May 2016, and had an additional 41 articles under consideration.
Figure~\ref{fig:article_stats} shows the monthly and cumulative publication rates;
on average, we published 8.5 articles per month, with some (nonstatistical) growth over time.


\begin{figure}[htbp]
    \centering
    \includegraphics[width=0.7\textwidth]{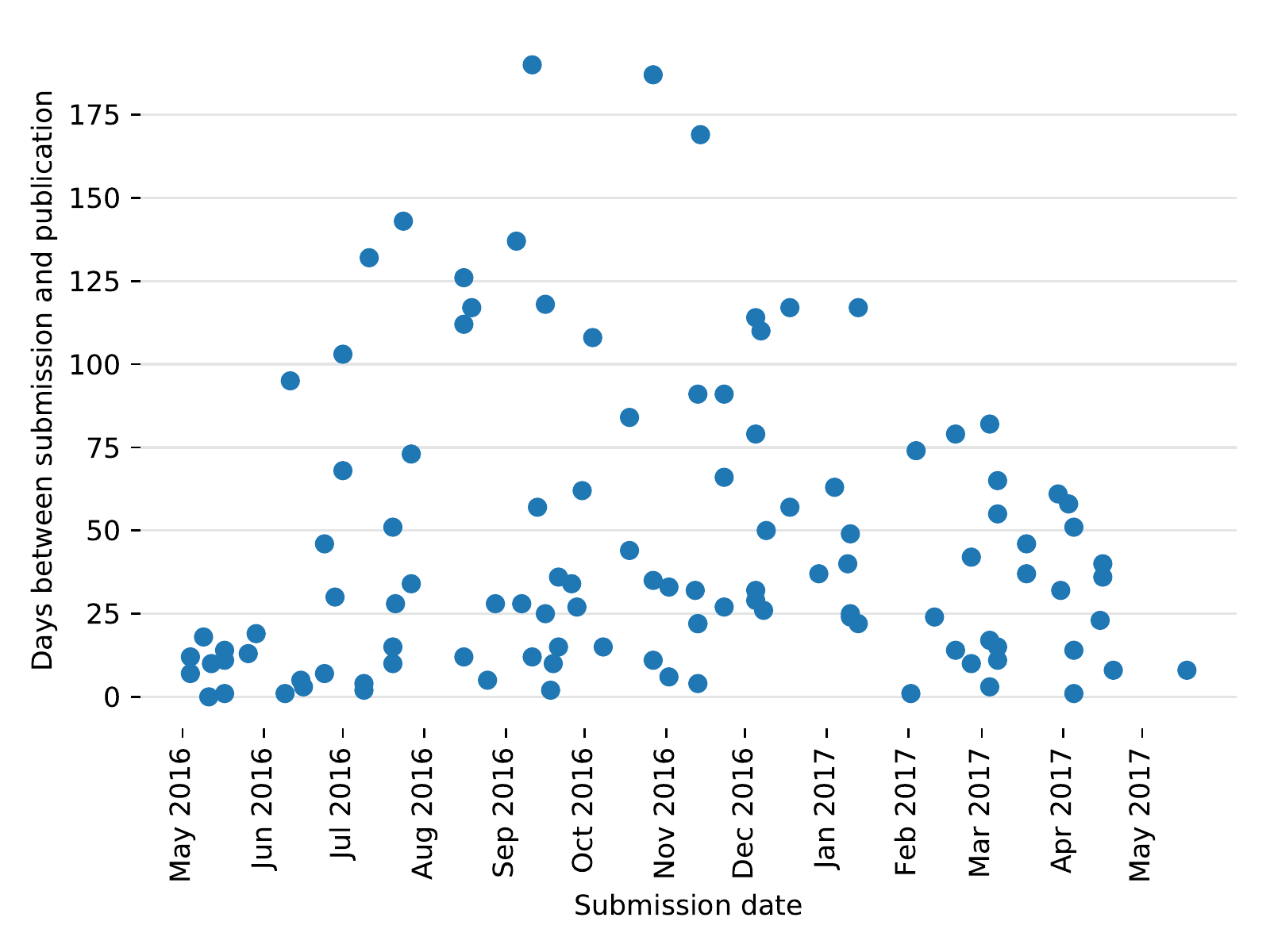}
    \caption{Days between submission and publication dates of the 111 articles \joss{} has published, between May 2016--May 2017.
    Data, plotting script, and figure file are available~\cite{JOSS-data-figs}.}
    \label{fig:article_review}
\end{figure}

Figure~\ref{fig:article_review} shows the numbers of days taken for processing
and review of the 111 published articles (i.e., time between submission
and publication), including finding a topic editor and reviewer(s). Since
the journal's inception in May 2016, articles spent on average 45.5 days
between submission and publication (median 32 days, interquartile range 52.3 days)
The shortest review took a single day, for \texttt{Application
Skeleton}~\cite{Zhang2016:joss}, while the longest review took 190 days,
for \texttt{walkr}~\cite{YuZhuYao2017:joss}. In the former case, the rapid turnaround
can be attributed to the relatively minor revisions needed (in addition to quick editor, reviewer, and author actions and
responses). In contrast, the latter case took much longer due to
delays in selecting an editor and finding an appropriate reviewer, and a
multimonth delay between selecting a reviewer and receiving reviews.
In other cases with long review periods, some delays in responding to requests
for updates may be attributed to reviewers (or editors) missing GitHub notifications
from the review issue comments.
We have already taken steps to improve the ability of authors, reviewers, and editors to keep track of their submissions, including a prompt to new reviewers to unsubscribe from the main \texttt{joss-reviews} repository~\cite{joss-reviews} (to reduce unnecessary notifications) and a weekly digest email for \joss{} editors to keep track of their submissions. In the future we may collect the email addresses of reviewers so we can extend this functionality to them.

\begin{figure}[htbp]
    \centering
    \includegraphics[width=0.7\textwidth]{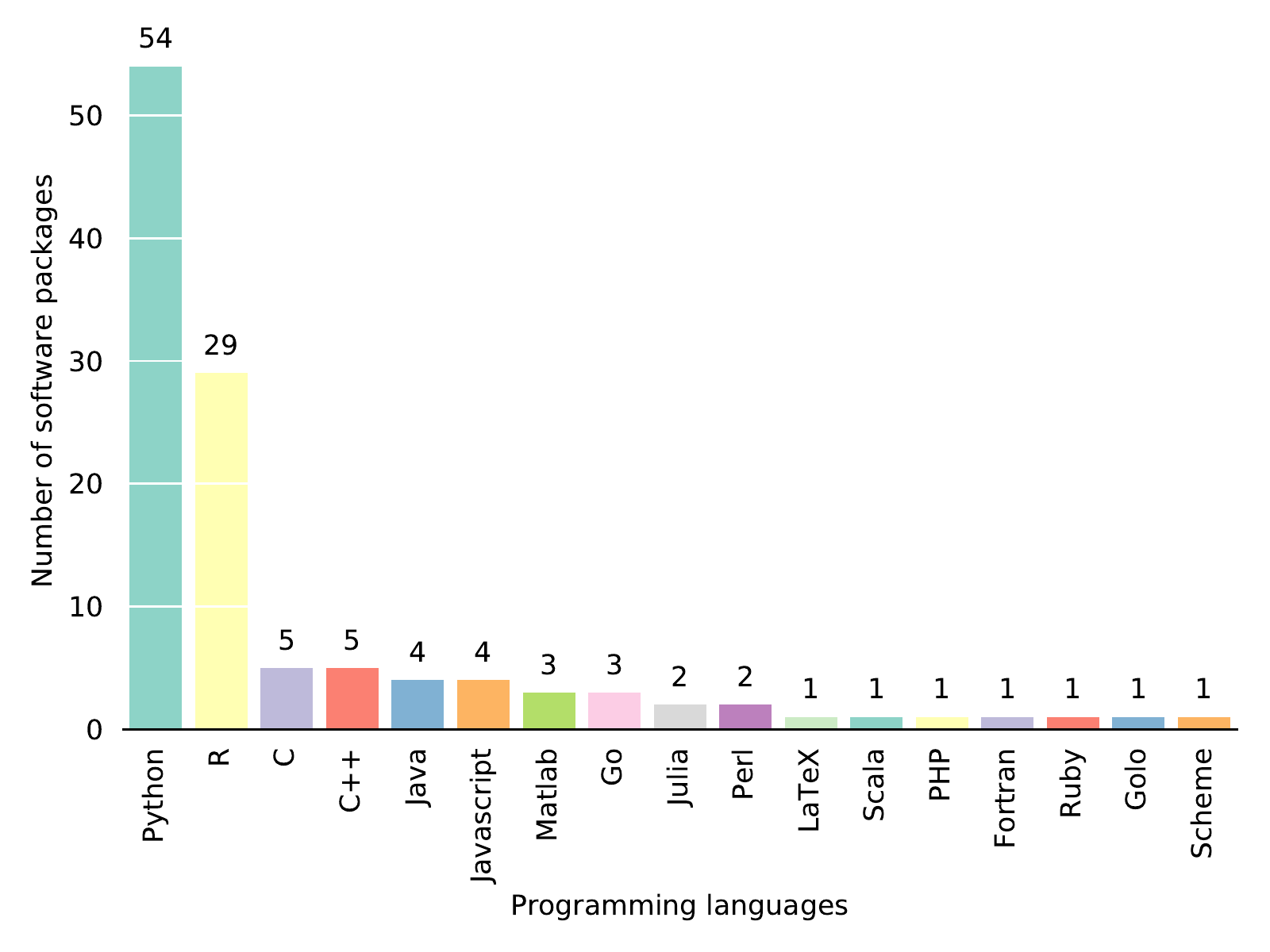}
    \caption{Frequency of programming languages from the software packages
    described by the 111 articles \joss{} published in its first year.
    Total sums to greater than 111, because some packages are multi-language. Data, plotting script, and figure file are available~\cite{JOSS-data-figs}.
    }
    \label{fig:programming_languages}
\end{figure}

Figure~\ref{fig:programming_languages} shows the frequency of programming languages appearing
in \joss{} articles. Python appears the most with over half of published software articles
(54), while R is used in nearly one-third of articles (29).
We believe the popularity of Python and R in \joss{} submissions is the result of (1) the adoption of these languages (and open-source practices) in scientific computing communities and (2) our relationship with the rOpenSci project.

Each article considered by \joss{} undergoes review by one
or more reviewers. The set of 111 published articles have been
reviewed by 93 unique reviewers.
The majority of articles received a review by
one reviewer (average of $1.11\pm 0.34$), with a maximum of
three reviewers.
Based on available data in the review issues, on average, editors reached out
to 1.85$\pm$1.40 potential reviewers (at most 8 in one case) via mentions
in the GitHub review issue. This does not include external communication,
e.g., via email or Twitter. Overall, \joss{} editors contacted 1.65
potential reviewers for each actual review (based on means).

Interestingly, the current reviewer list contains only 52 entries, as of
this writing~\cite{JOSS-reviewers}. Considering the unique reviewer count of 93,
we clearly have reached beyond those that volunteered to review a priori.
Benefits of using GitHub's issue infrastructure and our open reviews include:
1) the ability to tag multiple people, via their GitHub handles, to invite them as potential reviewers;
2) the discoverability of the work so that people may volunteer to review without being formally contacted;
3) the ability to get additional, unprompted feedback and comments; and
4) the ability to find reviewers by openly advertising, e.g., on social media.
Furthermore, GitHub is a well-known, commonly used platform where many (if not most) potential authors and reviewers already have accounts.

\begin{figure}[htbp]
    \centering
    \includegraphics[width=0.7\textwidth]{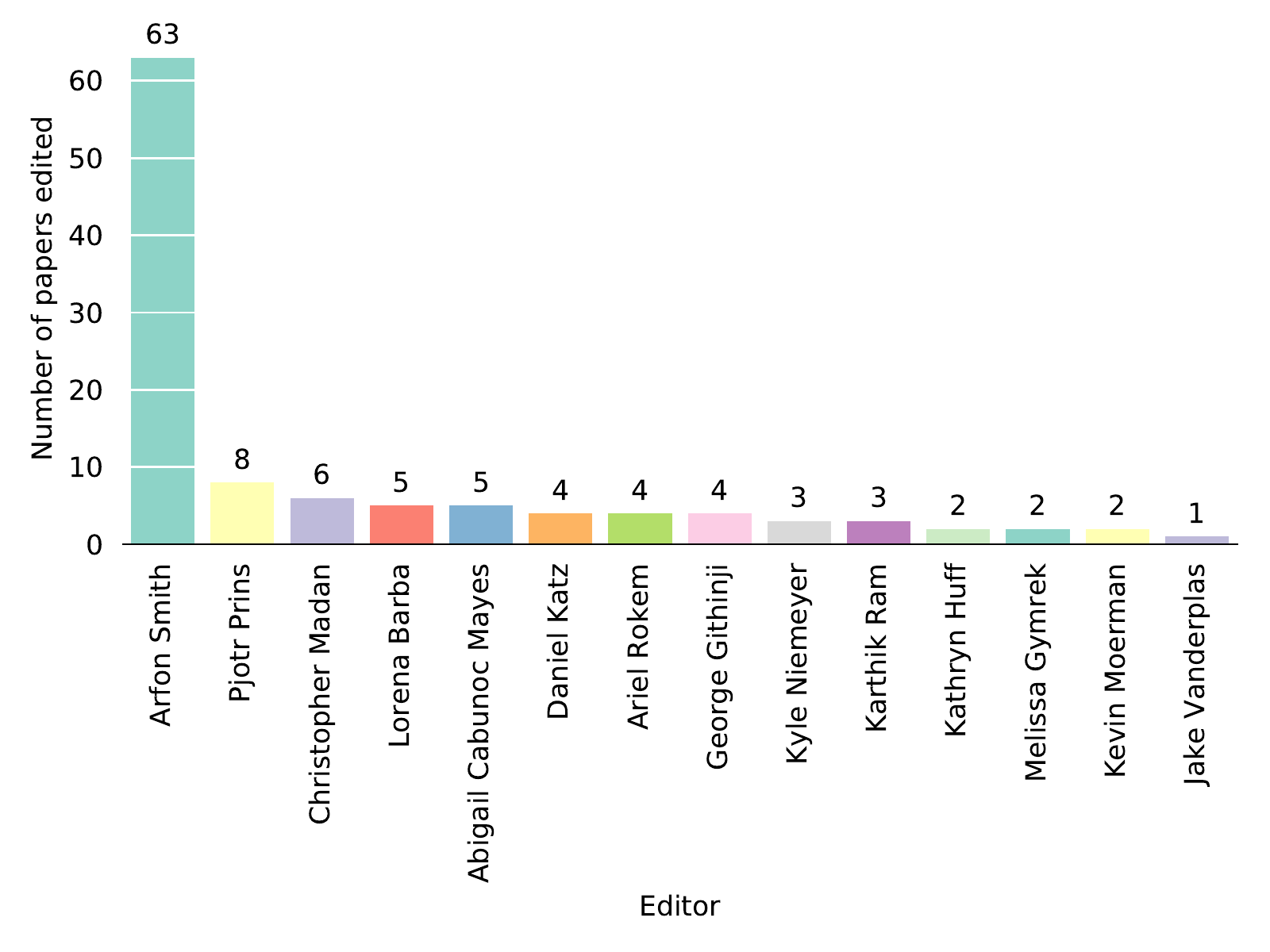}
    \caption{Numbers of articles handled by each of the \joss{}
    editors. Data, plotting script, and figure file are available~\cite{JOSS-data-figs}.
    }
    \label{fig:editors}
\end{figure}

Figure~\ref{fig:editors} shows the numbers of articles managed by each of the
\joss{} editors.
Editor-in-chief Arfon Smith stewarded
the majority of articles published in the
first year. This was somewhat unavoidable
in the first three months after launch, as
Smith served as the de facto sole editor for all submissions, with other members of the editorial
board assisting.
This strategy was not sustainable and, over time, we adopted the pre-review\slash review procedure to hand off articles to editors.
Also, authors can now select during submission the appropriate editor based on article topic.

\begin{figure}[htbp]
    \centering
    \includegraphics[width=0.7\textwidth]{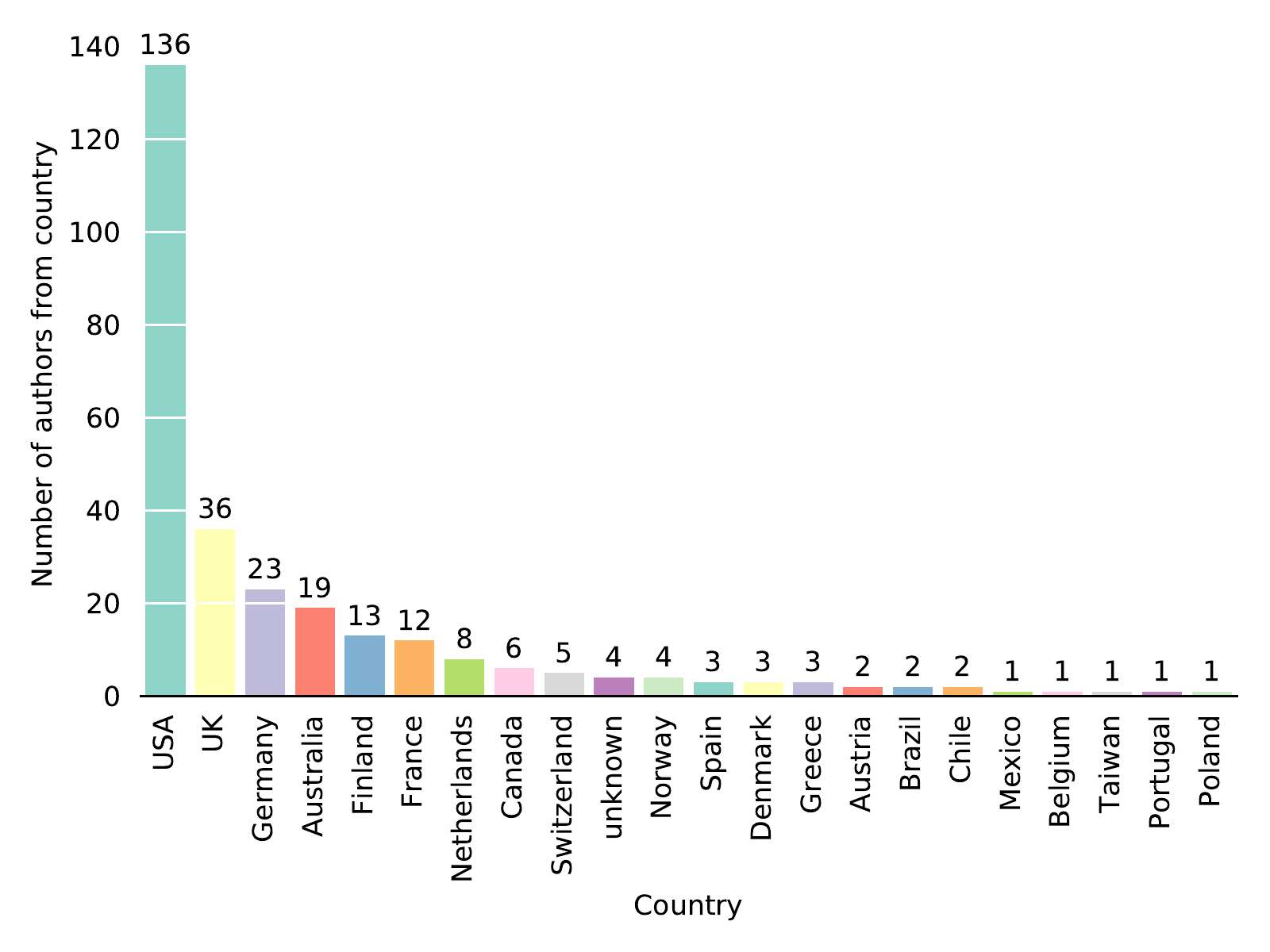}
    \caption{Numbers of authors from a particular country.
    Data, plotting script, and figure file are available~\cite{JOSS-data-figs}.
    }
    \label{fig:affiliations}
\end{figure}

Lastly, we analyzed the affiliations of the 286 authors associated with articles published in the first year.
Figure~\ref{fig:affiliations} shows the number of authors per country; we represented authors with multiple affiliations in different countries using their first affiliation. Authors with no affiliation, or where we could not identify the country, are shown as ``unknown.''
From the articles published in the first year, approximately 48\% of authors live in the United States and approximately 40\% live in Europe (including Switzerland). The remaining 12\% come from the rest of the world, most notably Australia (6.6\%) and Canada (2.1\%).
Moving forward, we hope to receive submissions from authors in more countries that even better represent who develops research software around the world; one strategy to achieve this involves continuing to expand our editorial board.

In its first year, \joss{} also developed formal relationships with two US-based nonprofit organizations. In March 2017, \joss{} became a community affiliate of the Open Source Initiative (\href{https://opensource.org}{opensource.org}), the steward of the open-source definition, which promotes open-source software and educates about appropriate software licenses.
And, in April 2017, \joss{} became a fiscally sponsored project of NumFOCUS (\href{https://www.numfocus.org}{numfocus.org}), a 501(c)(3) charity that supports and promotes ``world-class, innovative, open source scientific computing.''
Being associated with these two prominent community organizations increases the trust of the community in our efforts.
Furthermore, as a NumFOCUS project, \joss{} will be able to raise funding to sustain its activities and grow.

\section{The second year for \joss{}}

Our focus for the second year will be on continuing to provide a high-quality experience
for submitting authors and reviewers, and making the best use of the editorial board.
In our first year, we progressed from a model where the editor-in-chief handled
most central functions to one with more distributed roles for the editors,
particularly that of ensuring that reviews are useful and timely.
Editors can now select and self-assign to submissions they want to manage, while the editor-in-chief only assigns the remaining submissions.
As \joss{} grows, the process of distributing functions across the editorial board will continue to evolve---and more editors may be needed.

In the second year, we plan to complete a number of high-priority improvements to the \joss{} toolchain.
Specifically, we plan on automating the final steps for accepting an article.
For example, generating Crossref metadata and compiling the article are both currently handled by the editor-in-chief on his local machine using the \texttt{Whedon} RubyGem library.
In the future, we would like authors and reviewers to be able to ask the \texttt{Whedon-API} bot to compile the paper for them, and other editors should be able to ask the bot to complete the submission of Crossref metadata on their behalf.
Other improvements are constantly under discussion on the \joss{} GitHub repository
(\href{https://github.com/openjournals/joss/issues}{github.com/openjournals/joss/issues}).
In fact, anyone is able to report bugs and suggest enhancements to the experience.
And, since the \joss{} tools are open source, we welcome contributions in the form of bug-fixes or enhancements via the usual pull-request protocols.

Beyond roles and responsibilities for the editors, and improvements to the \joss{} tools and infrastructure, we will take on the more tricky questions about publishing software, such as how to handle new software versions.
Unlike traditional research articles that remain static once published,
software usually changes over time, at least for maintenance and to avoid
software rot\slash collapse (where software stops working because
of changes in the environment, such as dependencies on
libraries or operating system).
Furthermore, because all potential uses of the software are not
known at the start of a project, the need or opportunity arises to add
features, improve performance, improve accuracy, etc.
After making one or more changes, software developers frequently update the software with a new version number.
Over time, the culmination of these changes may result in a major update to the software, and with many new contributors a new version might correspond to a new set of authors if the software is published.
However, this process may not translate clearly to \joss{}.
The editorial board will accept a new \joss{} article published with each major version or even a minor version if the changes seem significant enough to the editor and reviewer(s), but we do not yet know if this will satisfy the needs of both developers and users (corresponding to \joss{} authors and readers, respectively).


The discussion about new software versions also generally applies to software forks, where software is copied and, after some divergent development, a new software package emerges.
Similar to how we handle new software versions, the \joss{} editorial board will consider publication of an article describing a forked version of software if it includes substantial changes from a previously published version.
Authorship questions may be more challenging when dealing with forks compared with new versions, since forks can retain varying amounts of code from the original projects.
However, while a version control history generally makes it easy to suggest people who should be authors, deciding on authorship can be difficult and subjective, and is therefore ultimately project-dependent.
We prefer to leave authorship decisions to the projects, with discussion taking place as needed with reviewers and editors.

\section{Conclusions}
\label{conclusions}

Software today encapsulates---and generates---important research knowledge, yet
it has not entered the science publication ecosystem in a practical way.
This situation is costly for science, through the lack of career progression for
valuable personnel: research software developers.
We founded \joss{} in response to the acute need for an answer to this predicament.
\joss{} is a venue for authors who wish to receive constructive peer feedback,
publish, and collect citations for their research software.
By encouraging researchers to develop their software following best practices, and then share and publish it openly, \joss{} supports the broader open-science movement.
The number of submissions confirms the keen demand for this publishing mechanism:
more than 100 accepted articles in the first year and more than 40 others under review.
By the end of 2017, \joss{} has published nearly 200 articles.
Community members have also responded positively when asked to review submissions
in an open and non-traditional format, contributing useful reviews of the submitted software.

However, we are still overcoming initial hurdles to achieve our goals. \joss{} is
currently not fully indexed by Google Scholar, despite the fact that \joss{}
articles include adequate metadata and that we made an explicit request for
inclusion in March 2017 (see GitHub \href{https://github.com/openjournals/joss/issues/130}{issue \#130}).
Also, we may need to invest more effort into raising awareness of good practices
for citing \joss{} articles.
That said, we have some preliminary citation statistics: according to Google Scholar, \texttt{corner.py}~\cite{cornerpy} and \texttt{Armadillo}~\cite{armadillo} have been cited the most at 116 and 79 times, respectively.
Crossref's Cited-by service---which relies on publishers depositing reference information---reports 45 and 28 citations for the same articles~\cite{crossref-cited-by}.
While most other articles have received no citations to-date, a few have been cited between one and five times.
We have had at least two ``repeat'' submissions, i.e., submissions of a new version with major changes from a prior version.
Clementi et al.~\cite{CClementi2017} published \texttt{PyGBe-LSPR}, a new version that added substantially new features over the original \texttt{PyGBe} of Cooper et al.~\cite{DCooper2016}.
Similarly, the software published by Sandersen and Curtin~\cite{Sanderson2017} extended on (and cited) their earlier article~\cite{Sanderson2016}.

The journal cemented its position in the first year of operation, building trust within the community of open-source research-software developers and growing in name recognition.
It also earned weighty affiliations with OSI and NumFOCUS, the latter bringing the opportunity to raise funding for sustained operations.
Although publishing costs are low at \$3--6 per article, \joss{} does need funding, with the editor-in-chief having borne the expenses personally to pull off the journal launch.
Incorporating a small article charge (waived upon request) may be a route to allow authors to contribute to \joss{} in the future, but we have not yet decided on this change.
Under the NumFOCUS nonprofit umbrella, \joss{} is now eligible to seek grants for sustaining its future, engaging in new efforts like outreach, and improving its infrastructure and tooling.

Outreach to other communities still unaware of \joss{} is certainly part of our growth strategy.
Awareness of the journal so far has mostly spread through word-of-mouth and social networking~\cite{tauber-blog,titus-blog}, plus a couple of news articles~\cite{Nature:joss,SDtimes:joss}.
As of August 2017, \joss{} is also listed in the Directory of Open Access Journals (DOAJ) (\href{https://doaj.org/toc/2475-9066}{doaj.org/toc/2475-9066}).
We plan to present \joss{} at relevant domain conferences,
like we did at the 2017 SIAM Conference
on Computational Science \& Engineering~\cite{JOSS-CSE-poster} and the 16th Annual Scientific Computing with Python Conference (SciPy 2017).
We are also interested in partnering
with other domain journals that focus on (traditional) research articles.
In such partnerships,
traditional peer review of the research would be paired with peer review of the software, with \joss{} taking responsibility for the latter.

Finally, the infrastructure and tooling of \joss{} have unexpected added values: while developed to
support and streamline the \joss{} publication process, these open-source tools generalize to a lightweight journal-management system.
The \joss{} web application and submission tool, the \texttt{Whedon} RubyGem library, and the \texttt{Whedon-API} bot could be easily forked to create overlay journals for other content types (data sets, posters, figures, etc.).
The original artifacts could be archived on other
services such as Figshare, Zenodo, Dryad, arXiv, or engrXiv\slash AgriXiv\slash LawArXiv\slash PsyArXiv\slash SocArXiv\slash bioRxiv.
This presents manifold opportunities to expand the ways we assign career credit to the digital artifacts of research.
\joss{} was born to answer the needs of research software developers to thrive in the current merit traditions of science, but we may have come upon a generalizable formula for digital science.

\section*{Acknowledgements}

This work was supported in part by the Alfred P.\ Sloan Foundation.
Work by K.~E.~Niemeyer was supported in part by the National Science Foundation (No.\ ACI-1535065).
Work by P.~Prins was supported by the National Institute of Health (R01 GM123489, 2017--2022).
Work by K.~Ram was supported in part by The Leona M.\ and Harry B.~Helmsley Charitable Trust (No.\ 2016PG-BRI004).
Work by A.~Rokem was supported by the Gordon \& Betty Moore Foundation and
the Alfred P.~Sloan Foundation, and by grants from the Bill \& Melinda Gates
Foundation, the National Science Foundation (No.\ 1550224), and the National
Institute of Mental Health (No.\ 1R25MH112480).

\printbibliography

\end{document}